**A. A. Uleiev[1], A. G. Magner[1], S. P. Maydanyuk[2,3], A. Bonasera[4], H. Zheng[5,*], S. N. Fedotkin[1], A. I. Levon[6], U. V. Grygoriev[1], T. Depastas[4]**

[1] *Institute for Nuclear Research, National Academy of Sciences of Ukraine, Nuclear Theory Department, Kyiv, Ukraine*

[2] *Institute for Nuclear Research, National Academy of Sciences of Ukraine, Nuclear Processes Theory Department, Kyiv, Ukraine*

[3] *Institute of Modern Physics, Chinese Academy of Sciences, Huizhou, China*

[4] *Cyclotron Institute, Texas A&M University, College Station, Texas, USA*

[5] *School of Physics and Information Technology, Shaanxi Normal University, Xi'an, China*

[6] *Institute for Nuclear Research, National Academy of Sciences of Ukraine, Heavy Ions Physics Department, Kyiv, Ukraine*

*Corresponding author: zhengh@snnu.edu.cn

# MACROSCOPIC APPROACHES TO ROTATING NEUTRON STARS[a]

The macroscopic model for a neutron star (NS) as a perfect liquid drop at equilibrium is extended to rotating systems with a small frequency $\omega$ within the effective-surface (ES) approach. The gradient surface terms of the NS energy density $\mathcal{E}(\rho)$ in the Equation of State are taken into account along with the volume components at the leading order over the leptodermic parameter $a/R \ll 1$, where $a$ is the ES crust thickness and $R$ is the mean NS radius. The macroscopic NS angular momentum at small frequencies $\omega$ is used for calculations of the adiabatic moment of inertia (MI) within the Kerr metric approach in the outer Boyer-Lindquist and inner Hogan coordinate forms. The NS MI, $\Theta = \widetilde{\Theta}/(1 - \mathcal{G}_{t\varphi})$, was obtained in terms of the statistically averaged MI, $\widetilde{\Theta}$, and its time and azimuthal-angle correlation, $\mathcal{G}_{t\varphi}$, as the sums of volume and surface components. The MI $\Theta$ depends dramatically on the effective radius $R$ due to strong gravitation and surface effects. We found significant additional rotational constraints on the radius $R$ due to the correlation term $\mathcal{G}_{t\varphi}$ and surface contributions. With these contributions, the adiabaticity condition is better fulfilled for a stronger gravitation in many well-known neutron stars.

---

[a] Presented at the XXXII Annual Scientific Conference of the Institute for Nuclear Research of the National Academy of Sciences of Ukraine, Kyiv, May 26 - 30, 2025.

## 1. Introduction

In this paper, in line with the derivation of Tolman-Oppenheimer-Volkoff (TOV) equations [1-3] from the General Relativistic Theory (GRT), we present the macroscopic model for neutron stars (NSs) as perfect liquid drops at equilibrium under a strong gravitation. They are rotating with a small azimuthal frequency $\omega$ around the symmetry axis at zero temperature [4,5]. In accordance with numerous recent observational data [6-20], the NS masses $M$ are larger than the Solar mass, $M \sim (1-2) M_{\odot}$, but their radii $R$ are extremely small, $R \sim 10$ km, such that the mean inner density $\overline{\rho}$ is larger than or of the order of $\rho_0$, where $\rho_0 \sim 10^{14}$ g/cm$^3$ is the nuclear matter density; see also early successfully obtained observational data for radio pulsars [21]. These properties mainly create a new field in astrophysics, i.e., nuclear astrophysics [22-34].

As the TOV equations [2] are used along with the Equation of State (EoS) in a lot of astrophysical works on NSs, it would be logical to agree the arguments for the specific derivations of the TOV equations and EoS. This is the main motivation of the macroscopic approach [35], which is also important for studying the NS rotations. Within nuclear astrophysics, we have to take into account the strong gravitational forces through the fundamental work [36] as in the derivations of the TOV equations, in contrast to nuclear physics. Our suggestion for these coordinates [37] does not exclude microscopic EoSs. However, it requires the corresponding essential modification of the TOV equations; see Refs. [35,37]. Taking Tolman's ideas, the EoS was extended [35,37] to that for a dense macroscopic system of particles for the case of non-rotating NS systems; see also Refs. [4,5]. In such a leptodermic system, one finds the density $\rho$ as function of the radial coordinate with exponentially decreasing behavior from an almost constant value $\overline{\rho}$ inside the dense system to zero through the NS effective surface (ES) in a relatively small (inner) crust range $a$, $a \ll R$. The ES is defined as the set of spatial coordinates with maximum density gradients. To obtain analytical solutions for the density distribution and EoS, we use the leptodermic approximation, $a/R \ll 1$ [28, 29, 32, 34, 38-46]. This imposes the limitations to the applicability for our macroscopic approach by sufficiently large effective radii $R$ of NSs for their masses, which are larger than or of the order of the Sun mass, and densities $\overline{\rho}$ larger than or of the order of that of nuclear matter.

In our macroscopic approach [35], the NS radius $R$ is the curvature radius of the NS ES. Within this ES approximation, the simple and accurate solutions of many nuclear and dense liquid-drop problems involving the density distributions were obtained for nuclei [47-52] and neutron stars [35, 37]. The ES approach exploits the property of saturation of the statistically averaged density $\rho$ inside of such a dense molecular- or nuclear-type and gravitational system, which is its characteristic macroscopic feature. Note that for the dense molecular (e.g., liquid-drop) systems, van der Waals (vdW) [53] suggested the phenomenological capillary theory; see also Refs. [5, 54]. This theory predicts the results for the density $\rho$ and surface tension coefficient $\sigma$. They are similar to the results obtained later in Ref. [47] for liquid drops, nuclei, and presumably, NSs [35]. The realistic energy-density distribution is minimal at a certain saturation density for a dense particle system in the infinite nuclear matter. As a result, a relatively narrow edge region exists in finite nuclei or NS crust in which the macroscopic density $\rho$ drops sharply from its almost central value to zero. We assume here that the statistically averaged density inside the system far from the ES can be relatively changed a little. This saturation property of the macroscopic dense system, such as a hydrostatic (hydrodynamic) liquid drop, e.g., nucleus, or probably, NS in a final evolution state, is considered macroscopically, according to the TOV derivations [1-3]. Thus, an easy extraction of relatively large terms in the density distribution equations for the variation equilibrium condition can be realized inside and near the ES. The equilibrium condition means that the variation of the total energy $E$ over the density $\rho$ is zero under the constraints that fix some integrals of motion beyond the energy $E$ by the Lagrange method. The Lagrange multipliers are determined by these constraints within the local energy-density theory, in particular, the extended Thomas-Fermi (ETF) approach, well known from nuclear physics, Refs. [55] and [56] (chapt. 4). The Lagrange equilibrium equations can be reduced to a simple one-dimensional catastrophe equation for the density $\rho$ in the leading normal-to-ES direction.

Such an equation mainly determines approximately the density distribution across the diffused surface layer of its thickness parameter $a$ to the mean ES-curvature radius $R$ ratio, $a/R \ll 1$, in the body-fixed coordinate system, i.e., with zero rotational frequency, $\omega = 0$, where $\omega = \mathrm{d}\varphi/\mathrm{d}t$; see Ref. [57]. A small leptodermic parameter, $a/R$, of the expansion within the ES approximation can be used for analytically solving the variational problem for the minimum of the system energy with constraints for a fixed particle number, and other integrals of motion, such as the angular momentum, quadrupole deformation, etc. When this edge distribution of the density is known, the leading static and dynamic density distributions that correspond to the diffused surface conditions can be simply constructed. We should derive the equation for the ES coupled to the NS volume by boundary conditions [48, 58]. This ES approach is based on the catastrophe theory for solving differential equations with a small parameter of the order of $a/R$ as the coefficient in front of the higher-order derivatives in normal to the ES direction. A relatively large change of the density $\rho$ on a small distance $a$ with respect to the ES curvature radius $R$ takes place for the liquid-matter drop (nuclei, water drops, and presumably, neutron stars). Inside such dense systems, the density $\rho$ is changed slightly around a mean internal-density constant $\overline{\rho}$ relatively far from the ES. Therefore, one obtains essential effects on the surface capillary pressure as in the general vdW theory [53,54].

The accuracy of the ES approximation was checked in Ref. [49] for the nuclear physics problems by comparing the results with the existing nuclear theories like Hartree-Fock (HF) [59] and ETF [55,56], based on the Skyrme forces [55,59-67], but for the simplest case without spin-orbit and asymmetry terms of the energy density functional. The extension of the ES approach to the nuclear isotopic symmetry and spin-orbit interaction has been done in Refs. [50-52]. The Swiatecki derivative terms of the symmetry energy for heavy nuclei [68-75] were taken into account within the ES approach in Ref. [52]. The discussions of the progress in nuclear physics and astrophysics within the relativistic local density approach can be found in reviews Refs. [32,34,76]; see also Refs. [46,77,78]. The ES corrections to the TOV equations [2,3] for neutron stars have been derived in Ref. [37].

The macroscopic NS rotation problem can be formulated first for a small rotational-energy perturbation of the strongly gravitating spherically-symmetric background described by the Schwarzschild metric. The basic ideas for studying a slow-rotating spherical system in the GRT were largely suggested already in Ref. [57] by Lense and Thirring to use an extended Schwarzschild metric. The frequency $\omega$ dependence of the gravitational metric for a slow-rotating star was first more consequently obtained by Kerr within the GRT [79]. Simple, clear derivations of the Kerr metric approach were discussed [4,80,81] for a relatively slow rotation by accounting also for small (quadrupole or spheroidal) deformations of the gravitating system. The specific independent formulation based on the Lense-Thirring [57] Schwarzschild metric linear-perturbation approach was developed in Ref. [82]. The simplest form of the Kerr metric in terms of other more transparent variables was found for the region outside of the NS by Boyer and Lindquist in Ref. [83], and for the inner NS part by Hogan in Ref. [84]; see also Refs. [4,80,81,85]. Friedman and his collaborators have developed the GRT for the energy-momentum of uniformly rotating stars as a perfect dense stellar fluid with a constant angular velocity $\omega$ [86-89]; see also the earlier work by Boyer and Lindquist, Ref. [90], and Ref. [5]. This theory of the rotating neutron stars will be specified as a linear response of the Schwarzschild gravitational metric to their rotational energy. In the present work, we extend the ES approximation of Refs. [35, 37, 50-52], to the rotating neutron stars for their slow macroscopic motion with constant frequency $\omega$ within this linear perturbation theory. This report is based on the extended work, Ref. [91].

## 2. Basic theoretical points

We start with the remarkable solution for the gravitational metric $g_{\mu\nu}$ of the GRT equations obtained by Kerr [79]. In the form of Boyer-Lindquist [83] and Hogan [84], the time-length element $ds^2$ in the linear approximation over the rotation frequency parameter $\Omega$ can be presented in units

of $c = G = 1$ as

$$ds^2 = e^\nu dt^2 + 2\tau\Omega \sin^2\theta\, dt d\varphi - e^\lambda dr^2 - r^2 d\theta^2 - r^2 \sin^2\theta\, d\varphi^2. \quad (1)$$

Here, $\nu$ and $\lambda$ are the Schwarzschild metric parameters [3, 4, 36]. According to Refs. [83,84], for $\tau(r)$ one finds

$$\tau = 1 - \left(A - \frac{1}{2}\sqrt{1 - \frac{r^2}{R_S^2}}\right)^2 \quad \text{at} \quad r \leq R, \quad (2)$$

and

$$\tau = \frac{r_g}{r} \qquad \text{at} \qquad r > R, \quad (3)$$

where

$$A = \frac{3}{2}\left(1 - \frac{R^2}{R_S^2}\right)^{1/2}, \quad (4)$$

$R_S$ is the Schwarzschild radius; see Ref. [3],

$$R_S = \sqrt{\frac{3}{8\pi\overline{\rho}}}, \quad (5)$$

and $r_g$ is the gravitational radius,

$$r_g = 2M \quad (6)$$

$M$ is the NS mass. A smooth transition of the outer to the inner Schwarzschild metrics leads approximately to the boundary condition at the effective NS radius $R$ [3],

$$\frac{r_g}{R} = \frac{R^2}{R_S^2}. \quad (7)$$

The NS angular momentum $I$ can be calculated in terms of the energy-density tensor $T_\mu^\nu$; see Refs. [86-89],

$$I = \int T_\mu^\nu \phi^\mu \hat{n}_\nu \mathrm{d}\mathcal{V}, \qquad I \approx \omega\Theta \approx \Omega M, \quad (8)$$

where $\phi^\mu$ is the Killing vector, and $\hat{n}_\nu$ is the normal vector to the hypersphere, which is a boundary of the integration volume occupied by the gravitating masses. The relation between the angular momentum $I$ and Kerr parameter $\Omega$ is valid outside of the gravitating system, and $\Theta$ is the moment of inertia (MI). The energy-momentum tensor $T_\mu^\nu$ for the NS as a perfect liquid drop is given by Refs. [4,5]

$$T_\mu^\nu = \mathcal{E}(\rho) u_\mu u^\nu + \mathcal{P} g_\mu^\nu, \quad (9)$$

where $g_\mu^\nu$ is the gravitational metric [see Eq. (1)], $u^\mu \propto (1,0,0,\overline{\omega})$ is the four-velocity for the NS rotation with a dimensionless frequency (angular momentum) $\overline{\omega} = I/M^2$, $\mathcal{P}$ is the pressure. The energy density $\mathcal{E}(\rho)$ is the ETF macroscopic Equation of State [35]:

$$\mathcal{E}(\rho) = \mathcal{A}(\rho) + \mathcal{C}(\nabla\rho)^2, \tag{10}$$

$$\mathcal{A} = \rho + \varepsilon(\rho), \tag{11}$$

$$\varepsilon(\rho) = \frac{K}{m\overline{\rho}^2}\rho(\rho - \overline{\rho})^2. \tag{12}$$

In the last equation, $K$ is the incompressibility modulus including the mean molecular vdW, nuclear Skyrme, and gravitational fields, and $m$ is the effective test-particle mass. In Eq. (10), $\mathcal{C}$ is the macroscopic inter-particle interaction constant for the same fields, which determines the crust thickness, $a \propto (m\mathcal{C}\overline{\rho}/K)^{1/2}$; see Appendix A and Refs. [35,91].

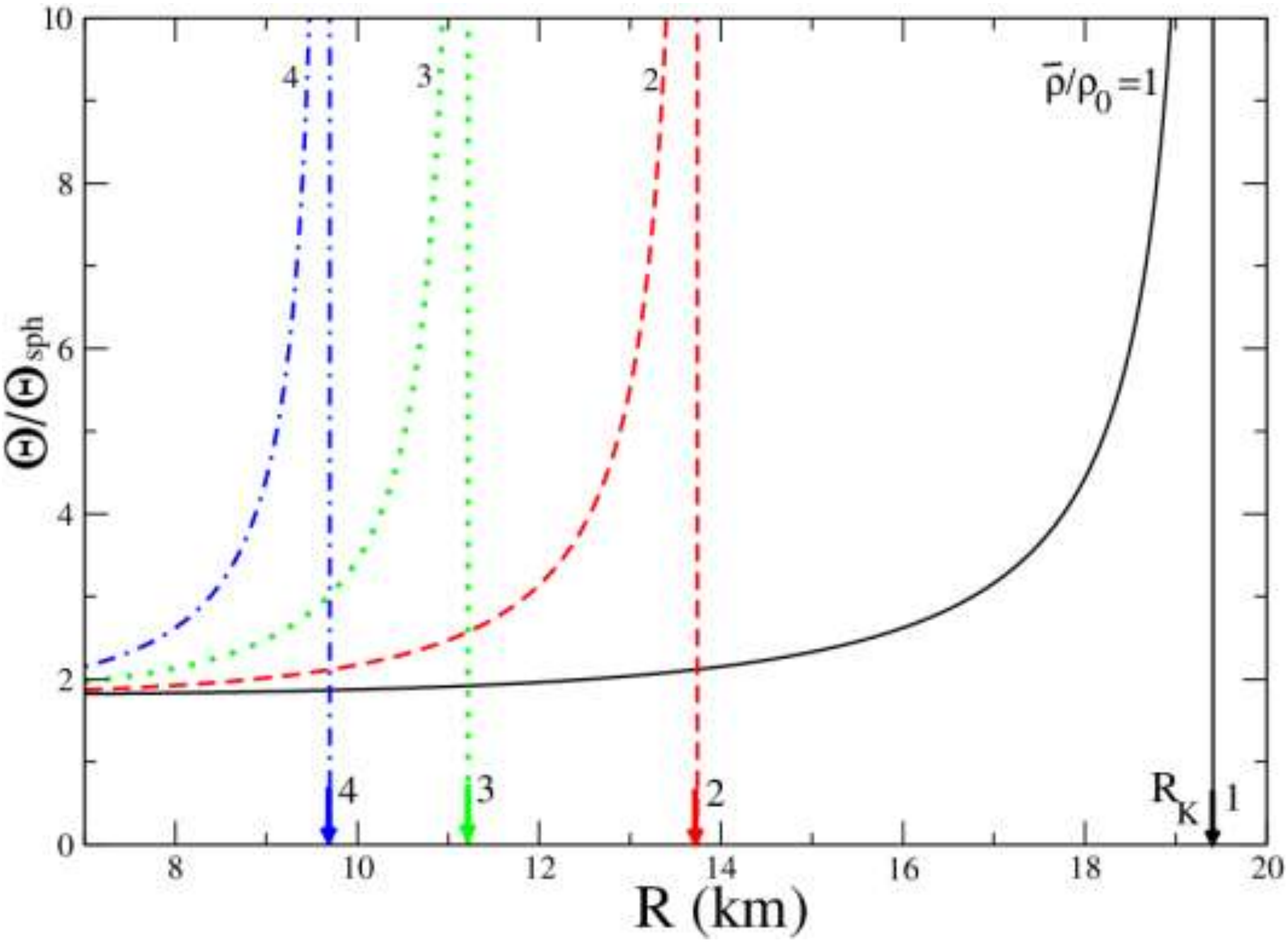

Fig. 1. Adiabatic moments of inertia $\mathbf{\Theta}$ in units of $\mathbf{2MR^2/5}$ as functions of the NS radius $\boldsymbol{R}$ are shown for inner densities, $\overline{\boldsymbol{\rho}}/\boldsymbol{\rho_0} = \mathbf{1-4}$, where $\boldsymbol{\rho_0} = \boldsymbol{m_N n_0} = \mathbf{2.68 \cdot 10^{14}\, g/cm^3}$ is the nuclear matter density. Arrows show the poles $\boldsymbol{R_K}$ of the MI, Eq. (14), for finite $\boldsymbol{t-\varphi}$ correlations $\boldsymbol{\mathcal{G}_{t\varphi}}$, Eq. (16).

In the adiabatic approximation for small rotational energy, one has

$$E_{rot} = \frac{1}{2}\Theta\omega^2 \ll E_{ETF} = E_V + E_S, \tag{13}$$

where $E_{ETF}$ is the total ETF energy, $E_V$ and $E_S$ are the volume and surface energy components on the very right; see Eq. (A10). For the moment of inertia (MI) $\Theta$, from Eqs. (8) and (9), one obtains

$$\Theta = \frac{\partial I}{\partial \omega} = \frac{\widetilde{\Theta}}{1 - \mathcal{G}_{t\varphi}}. \tag{14}$$

For the statistically averaged MI related to the diagonal Schwarzschild gravitational metric [2],

one has

$$\widetilde{\Theta} = \int \mathcal{E}(\rho) e^{-\nu} r^2 \sin^2\theta \,\mathrm{d}\mathcal{V}\,, \tag{15}$$

where $\mathrm{d}\mathcal{V} = J(r) r^2 \sin\theta \,\mathrm{d}r\mathrm{d}\theta\mathrm{d}\varphi$ (see Appendix A). The $t-\varphi$ correlation term in Eq. (1) leads to the following rotational contribution:

$$\mathcal{G}_{t\varphi} \approx \frac{2}{M} \int \mathcal{E}(\rho) e^{-\nu} \tau \sin^2\theta \,\mathrm{d}\mathcal{V}\,, \tag{16}$$

where $\tau$ is given by Eq. (2). This contribution appears in Eq. (14) due to a consistent accounting for the second relation between the angular momentum $I$ and the moment of inertia $\Theta$ in Eq. (8). Splitting now the energy density $\mathcal{E}(\rho)$, Eq. (10), into non-gradient, $\mathcal{A}(\rho)$, Eq. (11), and gradient, $\mathcal{C}(\nabla\rho)^2 \approx \mathcal{C}(\partial\rho/\partial r)^2$, dependent components, and using similar techniques as presented in Ref. [35], one can derive each of them, $\widetilde{\Theta}$ and $\mathcal{G}_{t\varphi}$, in terms of the volume and surface terms,

$$\widetilde{\Theta} = \widetilde{\Theta}_V + \widetilde{\Theta}_S, \qquad \mathcal{G}_{t\varphi} = \mathcal{G}_V + \mathcal{G}_S. \tag{17}$$

The specific expressions for the MI $\widetilde{\Theta}$, Eq. (15), and $\mathcal{G}_{t\varphi}$ contributions, Eq. (16), and their volume and surface components are shown in Appendix A.

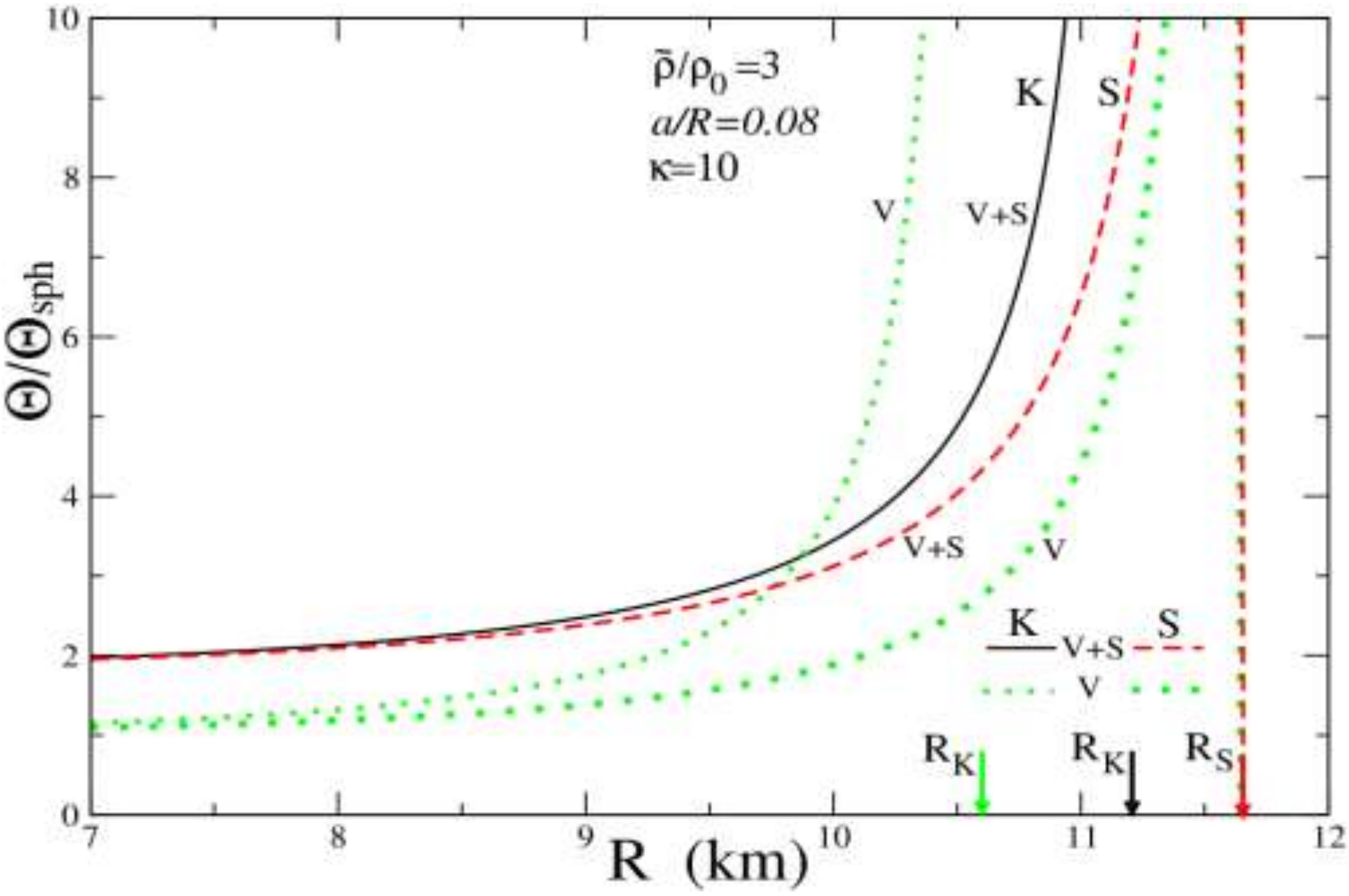

Fig. 2. The same as in Fig. 1 but for different approximations. Solid and dashed lines (V+S) show the MI contributions with and without the correlation term, respectively. Volume contributions are shown by frequent and rare dotted lines (V). The dimensionless incompressibility is given by $\boldsymbol{\kappa = 10}$; see Eq. (A14). The leptodermic parameter $\boldsymbol{a/R}$ is $\mathbf{0.08}$. Arrows show the Schwarzschild radius $\boldsymbol{R_S}$, Eq. (5), and rotational poles due to the $\boldsymbol{t-\varphi}$ correlations, $\boldsymbol{R_K}$.

For the adiabatic condition, Eq. (13), for the NS periods $P$, one finds

$$P \gg P_0, \qquad P_0 = 2\pi\sqrt{\Theta/2E_{ETF}} \tag{18}$$

For the volume contribution $P_0^{(V)}$ of the upper boundary period $P_0$, one obtains

$$P_0^{(V)} = 2\pi \sqrt{\frac{\widetilde{\Theta}_V}{2E_V(1-\mathcal{G}_V)}} \approx 2\pi R \sqrt{\frac{W_1\left({R}/{R_S}\right)}{1-W_2\left({R}/{R_S}\right)}}. \tag{19}$$

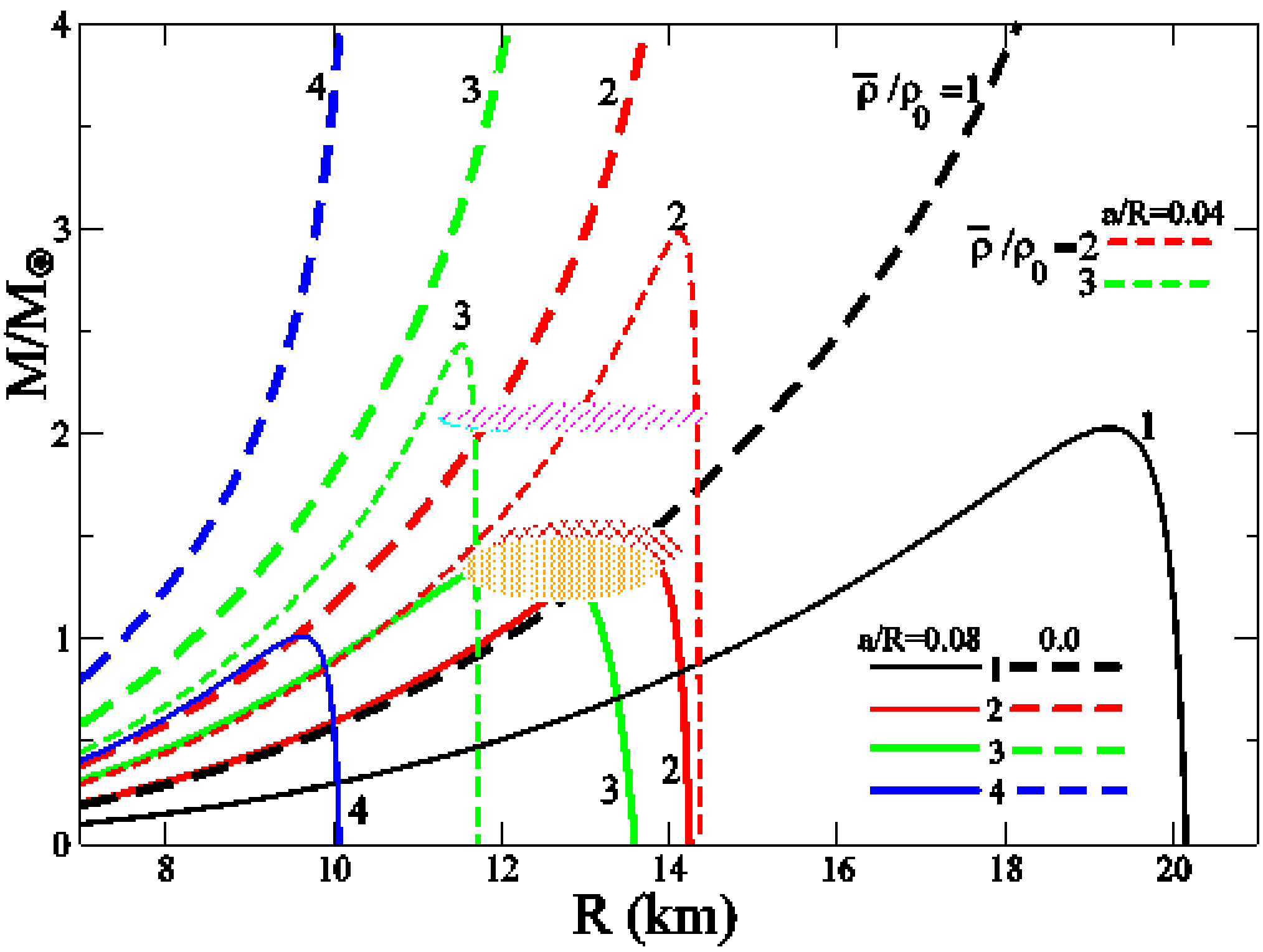

Fig. 3. NS masses $\boldsymbol{M}$, Eq. (A8), as functions of the NS radius $\boldsymbol{R}$ are shown by solid lines for relative densities $\overline{\boldsymbol{\rho}}/\boldsymbol{\rho_0} = \mathbf{1}-\mathbf{4}$, and leprodermic parameter $\boldsymbol{a}/\boldsymbol{R} = \mathbf{0.08}$. Dashed lines are the volume masses $\boldsymbol{M_V}$ (for $\boldsymbol{a}/\boldsymbol{R} = \mathbf{0}$). Frequent red and green dashed lines show masses for $\boldsymbol{a}/\boldsymbol{R} = \mathbf{0.04}$ and $\overline{\boldsymbol{\rho}}/\boldsymbol{\rho_0} = \mathbf{2}$ and $\mathbf{3}$. Red (oblique cells) and orange (frequent points) spots show observational data on the NS J0030+0441 [9,11]; magenta (oblique lines), and cyan (frequent points) spots present the J0740+6620 [10,19].

For the statistically averaged volume periods $\widetilde{P}_0^{(V)}$, one finds $(c = G = 1)$

$$\widetilde{P}_0^{(V)} = 2\pi \sqrt{\frac{\widetilde{\Theta}_V}{2E_V}} \approx 2\pi R \sqrt{W_1\left({R}/{R_S}\right)} \approx (0.11-0.27)\mathrm{ms}. \tag{20}$$

These estimates are obtained for typical NS masses $M = 0.6 - 2.5 M_{\odot}$, and radius $R = 10$ km, and approximately 0.15 – 0.24 ms for $R = 15$ km.

## 3. Discussions

Fig. 1 shows the adiabatic MI $\Theta$, Eqs. (14) - (16) (see Appendix A) as functions of the effective radius $R$ for different densities $\overline{\rho} = (1-4)\rho_0$. As seen from this Figure, the MI changes dramatically because of appearance of the pole [see Eq. (14)], $R = R_K$, due to the inclusion of the $t-\varphi$ correlation contribution $\mathcal{G}_{t\varphi}$, Eqs. (16) and (A5), (A18), and surface contributions. These rotational asymptotes are determined by the roots of the equation $\mathcal{G}_{t\varphi}(x) = 1$ with respect to the variable $x = R/R_S$. Therefore, one finds the additional constraints, $R < R_K$, because of strong gravitation in rotating NSs.

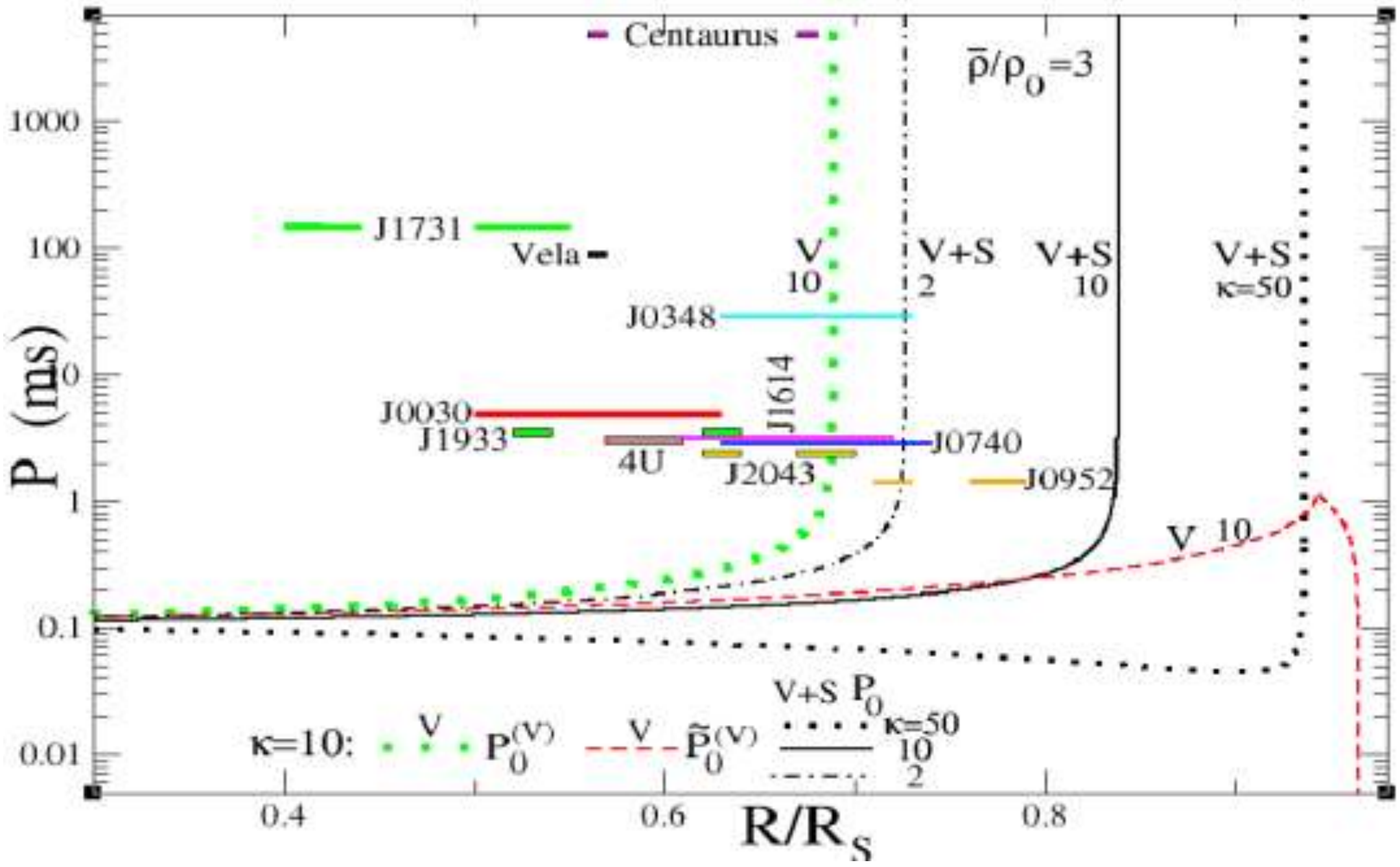

Fig. 4. NS periods $\boldsymbol{P}$ as functions of $\boldsymbol{R/R_S}$. Straight line segments represent observational data for several NSs; see Refs. [6-19]. Dashed-dotted, solid and dotted black lines are periods $\boldsymbol{P_0}$ at relative incompressibility values 2 (weak), 10 (strong) and 50 (super strong gravitation); see Eq. (A14) for $\boldsymbol{\kappa}$. Heavy green dotted and red dashed lines show the volume $\boldsymbol{P_0^{(V)}}$, Eq. (19), and $\boldsymbol{\tilde{P}_0^{(V)}}$, Eq. (20), without correlations. The parameters are $\boldsymbol{\bar{\rho}/\rho_0 = 3}$, $\boldsymbol{a/R = 0.08}$, and $\boldsymbol{R = 12}$ km.

Fig. 2 presents the same MI for the specific inner density, $\bar{\rho} = 3\rho_0$, but for different approaches. The solid and dashed lines show the MI with and without the correlation contribution $\mathcal{G}_{t\varphi}$, respectively. The frequent dotted and rare dotted lines display their volume contributions. As seen, the correlation contribution of $\mathcal{G}_{t\varphi}$ is significant, and the surface effects become larger when these correlations are included.

Fig. 3 shows the mass distribution $M(R)$, Eq. (A8), as a function of the effective radius $R$ by using only two physical parameters, the relative crust thickness $a/R$, and the relative asymptotical inner NS density, $\bar{\rho}/\rho_0 = 1-4$. The surface effect is measured by the relative difference between the full NS mass, $M = M_V + M_S$ (see Eq. (A8), solid lines) and its corresponding volume component $M_V$ ($a = 0$, Eq. (A9), dashed curves). As seen from Fig. 3, the surface component is quite significant even at a small leptodermic parameter $a/R = 0.08$ because of a strong gravitation. The NS mass $M(R)$ is not a monotonic function of $R$ for the fixed inner NS density, $\overline{\rho}$, because of the surface component, $M_S$. This contrasts with the monotonic behavior of the volume mass, $M_V \propto R^3$, in the Cartesian case of the weak Newtonian gravitational limit at $a = 0$. For any given value of $\overline{\rho}$, one finds a rather pronounced maximum in dependence on the full mass $M(R)$, Eq. (A8), MI $\Theta(R)$ (Figs. 1 and 2). The NS mass for each of these curves at a given value $\overline{\rho}$ disappears sharply in the limit $R \to R_S$, and does not exist at $R \geq R_S$. We should emphasize that our derivations for the surface components of the NS mass and MI fail near the point $R = R_S$. As seen from Fig. 3, our results are in reasonable agreement with the observational data, Refs. [10,8,9,14], at the volume density $\overline{\rho} = (2-3)\rho_0$, with $a/R = 0.08$ for the NS J0030+0441 [9,11], $M/M_\odot = 1.2 - 1.6$; and $a/R = 0.04$ for the NS J0740+6620 with a larger mass [10,19], $M/M_\odot = 2.0 - 2.1$.

Fig. 4 presents the characteristic upper limit periods $P_0$, Eq. (18), as a function of the ratio of the effective radius $R$ to the Schwarzschild radius $R_S$, Eq. (5), versus the results calculated for the observational data on the mass $M$, radius $R$, and period $P$. We show these data at the minimum and maximum values of $R/R_S$, which are expressed through the mass $M$ and radius $R$ by using the smooth joining boundary condition (7) for the outer and inner Schwarzschild metrics and the gravitational radius $r_g$, Eq. (6). The full characteristic rotation periods $P_0$ (solid) are compared with their volume $P_0^{(V)}$ (dotted) and the statistically averaged $\tilde{P}_0$ (dashed lines) contributions. Using the outer-inner boundary relation (7) and Eq. (6) for $r_g$, we expressed the variable $x = R/R_S$ in terms of the NS mass $M$ and radius $R$ by $x = (2GM/Rc^2)^{1/2}$ for a mean NS radius $R = 12$ km. The limit periods $P_0$ are dramatically changing functions of the relative effective NS radius, $R/R_S$ (for a given $R$), with the corresponding asymptotes $R = R_K$ for $P_0$ and $P_0^{(V)}$, and $R = R_S$ for $\tilde{P}_0^{(V)}$, respectively, as shown in Fig. 2. The periods $P_0$ and $P_0^{(V)}$ are monotonically increasing functions of $R/R_S$, which are asymptotically convergent to the corresponding asymptotes $R = R_K$. This contrasts with the dashed curve having a maximum for the statistically averaged period $\tilde{P}_0$, and a cusp in the top. In Fig. 4, we also show the essential dependence of the full limit periods $P_0$ on the dimensionless incompressibility parameter $\kappa$ [Eq. (A14)] through the tension coefficient $\sigma$, Eq. (A13), in the surface MI and energy components. The values of $P_0$ decrease with the incompressibility values of $\kappa$ from a weak ($\kappa = 2$, close to the nuclear matter case) to strong ($\kappa = 10$), and then, to super strong ($\kappa = 50$) gravity. As a result, for a strong and super strong gravitational field $\kappa = 10$ (solid) and 50 (frequent dotted black curve), the adiabatic condition is carried well for all stars shown in Fig. 4 while it is not the case for a weak gravitation ($\kappa = 2$). The adiabaticity condition is better fulfilled because of a stronger gravitation for several observational data (Fig. 4) for which the masses $M$, radii $R$, and rotational periods $\boldsymbol{P}$ are measured with good accuracy. The surface effect is increasing because of the correlation term and a change of the asymptote $R = R_K$, as seen from Figs. 2 and 4. Thus, we may conclude from the results shown in Fig. 4 that the adiabaticity condition (18) is carried out well for shown NSs at the strong $\kappa = 10$ and super strong gravitation $\kappa = 50$.

## 4. Conclusions

The macroscopic effective surface approximation is extended to a NS rotating dense-liquid drop at equilibrium for small frequencies $\omega$. The gravitational field was taken into account through the Kerr metric based on the Schwarzschild GRT solutions for spherical symmetry in the adiabatic approximation. The NS masses $M(R)$ ($r_g < R < R_S$) with surface components are in good agreement with recent observational data. The additional constraints on the NS radius $R$ appear due to the rotation and effective-surface contributions $r_g < R < R_K$, where $R_K$ is the root of equation $\mathcal{G}_{t\varphi}(x) = 1$. The adiabatic NS moment of inertia $\Theta$ depends much on the $t - \varphi$ correlation and surface contributions. The adiabatic condition is carried out well for many NSs with periods $5 - 3000$ ms, especially for a strong gravitation, even for smaller periods and larger masses.

For perspectives, one should systematically study our macroscopic method based on the adiabatic condition for the MI calculations in order to apply it to a wide range of neutron stars. Our macroscopic model should be improved by taking into account the internal many component structure, in particular, the isotopic asymmetry, and the thermal evolution. For including the gravitational-wave emission, one should essentially modify this model.

## Acknowledgments

The authors greatly acknowledge V. I. Abrosimov, C. A. Chin, O. Ya. Dzyublik, V. Z. Goldberg, M. I. Gorenstein, F. A. Ivanyuk, J. Holt, C. M. Ko, E. I. Koshchiy, J. B. Natowitz, S. A. Omelchenko, A. I. Sanzhur, G. V. Rogachev, Y. V. Shtanov, Yu. V. Taistra, S. I. Vacaru, V. I. Zhdanov for many creative and useful discussions.

## Appendix A

## Calculations of the moment of inertia

The statistically averaged $\widetilde{\Theta}$, Eq. (15), and the $t - \varphi$ correlation $\mathcal{G}_{t\varphi}$, Eq. (16), MI can be explicitly analytically calculated within the linear approximation over the dimensionless frequency $\overline{\omega} = Ic/M^2G \ll 1$. Within this Appendix, we will recover for convenience the ordinary dimensions of quantities. The volume components $\widetilde{\Theta}_V$ and $\mathcal{G}_V$ are obtained in the closed and simple forms. For $\widetilde{\Theta}_V$, one finds (see Ref. [91])

$$\widetilde{\Theta}_V = \frac{8\pi\overline{\rho}}{3}\int_0^R e^{-\nu} J(r) r^4 \mathrm{d}r = \frac{2}{5} M R^2 W_1(z_0), \qquad \nu = 2\ln\left(A - \frac{1}{2}\sqrt{1 - \frac{r^2}{R_S^2}}\right), \tag{A1}$$

where $\nu$ is the parameter of the inner part of the Schwarzschild metric, Eq. (1) (see Ref. [3]), $z_0 = (1 - R^2/R_S^2)^{1/2}$, $A$ is given by Eq. (4), $J(r) = (1 - r^2/R_S^2)^{-1/2}$ is the inner radial Jacobian $(r < R)$, $M$ is the NS mass, $R_S = [3c^2/(8\pi G\overline{\rho})]^{1/2}$ is the Schwarzschild radius (the gravitational radius is $r_g = 2M/c^2$). In Eq. (A1),

$$W_1(z_0) = \frac{5}{(1 - z_0^2)^{5/2}}[q_1(1) - q_1(z_0)], \tag{A2}$$

where $z = \sqrt{1 - r^2/R_S^2}$ for the transformation of the radial variable $r$ to dimensionless $z$,

$$q_1(z) = \frac{2\sqrt{1 - z^2}(2 - 24A^2 + 6Az + z^2)}{2A - z} + 6(8A^2 - 1)\arcsin(z) + \\ + 24A\sqrt{1 - 4A^2}\ln[\zeta(z, A)], \tag{A3}$$

$$\zeta(z, A) = \frac{1 - 2Az + \sqrt{1 - 4A^2}\sqrt{1 - z^2}}{2A - z}. \tag{A4}$$

For the volume $t - \varphi$ correlation contribution $\mathcal{G}_V$, Eq. (16), one obtains

$$\mathcal{G}_V = \frac{16\pi\overline{\rho}}{3M}\int_0^R e^{-\nu} J(r) \tau r^2 \mathrm{d}r \approx W_2(z_0). \tag{A5}$$

Here, $\tau(r)$ is given by Eq. (2),

$$W_2(z_0) = \frac{2}{(1 - z_0^2)^{3/2}}[q_2(1, A) - q_2(z_0, A)], \tag{A6}$$

with the same $z$ and $z_0$, and

$$q_2(z) = \frac{(z^2 - 2Az + 8)\sqrt{1 - z^2}}{2(2A - z)} - \frac{9}{2}\arcsin(z) + \frac{8A}{\sqrt{1 - 4A^2}}\ln[\zeta(z, A)]; \tag{A7}$$

see Eq. (A4) for $\zeta$.

For the NS mass $M$, one results in

$$M \approx M_V + M_S = M_V\left[1 - \frac{4aR^2}{R_S^3 f(R/R_S)(1 - R^2/R_S^2)^{1/2}}\right], \tag{A8}$$

where $M_V$ is the volume mass component,

$$M_V = 2\pi\overline{\rho}R_S^3 f(x), \qquad f(x) = \arcsin(x) - x\sqrt{1 - x^2}, \qquad x = R/R_S. \tag{A9}$$

Similarly, for the total NS energy $E_{ETF}$, one has [35]

$$E_{ETF} = \int \mathcal{E}(\rho) d\mathcal{V} \approx E_V + E_S, \tag{A10}$$

where $E_V$ is the volume part of the total energy,

$$E_V = 2\pi\overline{\rho}c^2 R_S^3 f(x), \tag{A11}$$

with the same $x$ as in Eq. (A9). For the surface part, one obtains

$$E_S \approx \sigma S, \tag{A12}$$

where $S = 4\pi R^2$ is the surface area value, and $\sigma$ is the leading-order tension coefficient,

$$\sigma \approx \frac{4aK\overline{n}}{135\sqrt{1 - R^2/R_S^2}} = \frac{16a\kappa\overline{\rho}c^2}{45\sqrt{1 - R^2/R_S^2}}. \tag{A13}$$

Here, $\overline{n}$ is the inner particle-number density, $\kappa$ is the dimensionless incompressibility,

$$\kappa = \frac{K\overline{n}}{12\overline{\rho}c^2} = \frac{K}{12mc^2}, \tag{A14}$$

$\overline{\rho}c^2$ is the inner energy density constant, and $\overline{\rho} = m\overline{n}$. The surface MI components, $\widetilde{\Theta}_S$ and $\mathcal{G}_S$, are derived analytically in terms of the surface tension coefficient $\sigma$ (see Eq. (A13)),

$$\widetilde{\Theta}_S = \frac{4\pi}{3c^2} \frac{\sigma R^4}{\sqrt{1 - R^2/R_S^2}} \tag{A15}$$

and

$$\mathcal{G}_S = \frac{8\pi\sigma}{3Mc^2} \frac{R^4/R_S^2}{1 - R^2/R_S^2}. \tag{A16}$$

Notice that $\widetilde{\Theta}_S > 0$ and $\mathcal{G}_S > 0$ because the tension coefficient $\sigma$ is positive and $R < R_S$ for a stable equilibrium. These surface components are proportional to the leptodermic parameter $a/R$, $\sigma \propto a/R$; see Eq. (A13). The crust thickness $a$ is related to the inter-particle interaction parameter $\mathcal{C}$ of Eq. (10), the particle number density $\overline{n}$, and the total nuclear-gravitational incompressibility $K$ of Eq. (12) by $a = (18m^2\mathcal{C}\overline{n}/K)^{1/2}$ (see Ref. [35]).

## REFERENCES

1. R. C. Tolman. Static solutions of Einstein's field equations for spheres of fluid. Phys. Rev. 55 (1939) 364. DOI: 10.1103/PhysRev.55.364.
2. J. R. Oppenheimer, G. M. Volkoff. On massive neutron cores. Phys. Rev. 55 (1939) 374. DOI: 10.1103/PhysRev.55.374.
3. R.C. Tolman. Relativity, Thermodynamics, and Cosmology. (Dover Publications, New York, 1987; Oxford, the University Press, 1934, 1946, 1949, 1987).
4. L.D. Landau and E.M. Lifshitz, Theoretical Physics, v.2 Field theory (Butterworth-Heinemann, New York, 2003; FIZMATLIT, Moscow, 2003).

5. L.D. Landau and E.M. Lifshitz, Theoretical Physics, v.6 Fluid mechanics (Pergamon Press, 1987, FIZMATLIT, Moscow, 2013).
6. J. Nättilä et al. Neutron star mass and radius measurements from atmospheric model fits to X-ray nurst cooling tail spectra. Astron. Astrophys. 608 (2017) A31.
DOI: 10.105/0004-5361/201731082.
7. B. P. Abbott et al. Properties of the binary neutron star merger GW170817. Phys. Rev. X 9 (2019) 011001.
DOI: 10.1103/PhysRevX.9.011001.
8. G. Raaijmakers et al. Constraints on the dense matter equation of state and neutron star properties from NICER's mass–radius estimate of PSR J0740+6620 and multimessenger observations. Astrophys. J. Lett. 918 (2021) L29.
DOI: 10.3847/2041-8213/ac089a.
9. T. E. Riley et al. A NICER view of PSR J0030+ 0451: millisecond pulsar parameter estimation. Astrophys. J. Lett. 887 (2019) L21 (9 pp.).
DOI: 10.3847/2041-8213/ab481c.
10. T. E. Riley et al. A NICER view of the massive pulsar PSR J0740+ 6620 informed by radio timing and XMM-Newton spectroscopy. Astrophys. J. Lett. 918 (2021) L27.
DOI: 10.3847/2041-8213/ac0a81.
11. M. C. Miller et al. PSR J0030+ 0451 mass and radius from NICER data and implications for the properties of neutron star matter. Astrophys. J. Lett. 887.1 (2019) L24 ; Astrophys. J. 888 (2019) 12.
DOI: 10.48550/arXiv.1912.05705.
12. M. C. Miller et al. The radius of PSR J0740+ 6620 from NICER and XMM-Newton data. Astrophys. J. Lett. 918 (2021) L28.
DOI 10.3847/2041-8213/ac089b.
13. P. T. H. Pang et al., nuclear physics multimessenger astrophysics constraints on the neutron star equation of state: adding NICER's PSR J0740+ 6620 measurement. Astrophys. J. Lett. 922 (2021) 14.
DOI 10.3847/1538-4357/ac19ab.
14. V. Doroshenko at al. A strangely light neutron star. Nature Astronomy 6 (2022) 1444.
DOI: 10.21203/rs.3.rs-1509469/v1.
15. G.G.L. Nashed. Confront $f(R,T) = \mathcal{R} + \beta T$ modified gravity with the massive pulsar PSRJ0740+6620. arXiv:2308.08565 [gr-qc] (2023).
16. R. Kumar et al. Observational constraint from the heaviest pulsar PSR J0952-0607 on the equation of state of dense matter in relativistic mean field model. arXiv:2306.05097v1 (2023).
17. F. Xie et al. First detection of polarization in X-rays for PSR B0540-69 and its nebula. Astrophys. J. 962 (2024) 92.
DOI: 10.3847/1538-4357/ad17ba.
18. D. Sen and A. Guha. Estimating the dark matter halo velocity and surface temperature of some known

pulsars due to dark matter capture. arXiv:2402.13795 [hep-ph] (2024).

19. A. J. Dittmann et al. A more precise measurement of the radius of PSR J0740+ 6620 using updated NICER data. Astrophys. J. 974 (2024) 295.
DOI: 10.3847/1538-4357/ad5f1e.
20. Y. Kini et al. Constraining the properties of the thermonuclear burst oscillation source XTE J1814-338 through pulse profile modelling. Mon. Not. Roy. Astron.Soc. 535 (2024) 1507.
DOI: 10.1093/mnras/stae2398.
21. T. M. Tauris et al. Formation of double neutron star systems. Astrophys. J. 846 (2017) 170.
DOI: 10.3847/1538-4357/aa7e89.
22. G. Baym, H. A. Bethe, C. J. Pethick. Neutron star matter. Nucl. Phys. A 175 (1971) 225.
DOI: 10.1016/0375-9474(71)90281-8.
23. R. B. Wiringa, V. Fiks, A. Fabrocini. Equation of state for dense nucleon matter. Phys. Rev. C 38 (1988) 1010.
DOI: 10.1103/PhysRevC.38.1010.
24. B.-A. Li, L.-W. Chen, C. M. Ko. Recent progress and new challenges in isospin physics with heavy-ion reactions. Phys. Rep. 464 (2008) 113.
DOI: 10.1016/j.physrep.2008.04.005.
25. H. Zheng, A. Bonasera. Non-Abelian behavior of α bosons in cold symmetric nuclear matter. Phys. Rev. C 83 (2011) 057602.
DOI: 10.1103/PhysRevC.83.057602.
26. S. Gandolfi et al. The equation of state of neutron matter, symmetry energy and neutron star structure. Eur. Phys. J. A 50 (2014) 10.
DOI: 10.1140/epja/i2014-14010-5.
27. H. Zheng, G. Giuliani, A. Bonasera. Coulomb corrections to the extraction of the density and temperature in non-relativistic heavy ion collisions. J. Phys. G: Nucl. Part. Phys. 41 (2014) 055109.
DOI: 10.1088/0954-3899/41/5/055109.
28. A. F. Fantina et al. Neutron star properties with unified equations of state of dense matter. Astron. Astrophys. 559 (2013) A128.
DOI: 10.1051/0004-6361/201321642.
29. Y. Potekhin et al. Analytical representations of unified equations of state for neutron-star matter. Astron. Astrophys. 560 (2013) A48.
DOI: 10.1051/0004-6361/201321697.
30. A. Bauswein, S. Goriely, H.-T. Janka. Systematics of dynamical mass ejection, nucleosynthesis, and radioactively powered electromagnetic signals from neutron-star mergers. Astrophys J. 773 (2013) 78.
DOI: 10.1088/0004-637X/773/1/78.
31. A.F. Fantina et al. Stellar electron-capture rates on nuclei based on Skyrme functionals. EPJ Web of Conferences. Vol. 66. EDP Sciences, 2014.

DOI: 10.1051/epjconf/20146602035.

32. G. Baym et al. From hadrons to quarks in neutron stars: a review. Rep. Prog. Phys. 81 (2018) 056902.
DOI: 10.1088/1361-6633/aaaea6.
33. S. Goriely. Nuclear properties for nuclear astrophysics studies. Eur. Phys. J. A 59 (2023) 16.
DOI: 10.1140/epja/s10050-023-00973-1.
34. B. Sun, S. Bhattiprolu, J. M. Lattimer. Compiled properties of nucleonic matter and nuclear and neutron star models from nonrelativistic and relativistic interactions. Phys. Rev. C 109 (2024) 055801.
DOI: 10.1103/PhysRevC.109.055801.
35. A. G. Magner et al. Symmetry energy and neutron skin from surface properties of finite nuclei. Int. J. Mod. Phys. E 33 (2024) 2450043.
DOI: 10.1142/S0218301324500435.
36. K. Schwarzschild. Über das Gravitationsfeld eines Massenpunktes nach der Einsteinschen Theorie. Sitzungsber. K. Preuss. Akad. Wiss. (1916) 424.
37. A. G. Magner et al. Leptodermic corrections to the TOV equations and nuclear astrophysics within the effective surface approximation. arXiv:2409.04745 [nucl-th] (2024); accepted by the Nucl. Phys. A 1064 (2025) 123239.
DOI: 10.1016/j.nuclphysa.2025.123239.
38. J.M. Lattimer, M. Prakash. Equation of state, neutron stars and exotic phases. Nucl. Phys. A 777 (2006) 479.
DOI: 10.1016/j.nuclphysa.2005.01.014.
39. C.J. Horowitz, J. Piekarewicz. Neutron star structure and the neutron radius of 208Pb. Phys. Rev. Lett. 86 (2001) 5647.
DOI: 10.1103/PhysRevLett.86.5647.
40. C.J. Horowitz and J. Piekarewicz. Neutron radii of $^{208}$Pb and neutron stars. Phys. Rev. C 64 (2001) 062802.
DOI: 10.1103/PhysRevC.64.062802.
41. A.F. Fantina, F. Gulminelli. Nuclear physics inputs for dense-matter modelling in neutron stars. J. Phys.: Conf. Ser. 2586 (2023) 012112.
DOI: 10.1088/1742-6596/2586/1/012112.
42. H. Dinh Thi, A.F. Fantina, F. Gulminelli. Light clusters in the liquid proto-neutron star inner crust. Eur. Phys. J. A 59 (2023) 292.
DOI: 10.1140/epja/s10050-023-01199-x.
43. P. Haensel, A.Y. Potekhin, D.G. Yakovlev. Neutron Stars 1. Equation of State and Structure. Astrophys. Space Sci. Libr., vol. 326 (Springer, New York, 2007).
44. Y. Lim, J.W. Holt. Bayesian modeling of the nuclear equation of state for neutron star tidal deformabilities and GW170817. Eur. Phys. J. A 55 (2019) 209.
DOI: 10.1140/epja/i2019-12917-9.

45. S.L. Shapiro, S.A. Teukolsky. Black Holes, White Dwarfs, and Neutron Stars: The Physics of Compact Objects (Wiley-VCH, Weinheim, 2004).
DOI: 10.1002/9783527617661.
46. N.N. Shchechilin et al. Unified equations of state for cold nonaccreting neutron stars with Brussels-Montreal functionals. IV. Role of the symmetry energy in pasta phases. Phys. Rev. C 108 (2023) 025805.
DOI: 10.1103/PhysRevC.108.025805.
47. V.M. Strutinsky, A.S. Tyapin. Quasistatic Drop Model of the Nucleus as an Approximation to the Statistical Model. Exp. Theor. Phys. (USSR) 18 (1964) 664.
DOI: 10.1142/9789812815941_0018.
48. V.M. Strutinsky, A.G. Magner, M. Brack. The nuclear surface as a collective variable. Z. Phys. A 319 (1984) 205.
DOI: 10.1007/BF01415634.
49. V.M. Strutinsky, A.G. Magner, V.Yu. Denisov. Density distributions in nuclei. Z. Phys. A 322 (1985) 149.
DOI: 10.1007/BF01412028.
50. A.G. Magner, A.I. Sanzhur, A.M. Gzhebinsky. Asymmetry and spin-orbit effects… Int. J. Mod. Phys. E 18 (2009) 885.
DOI: 10.1142/S0218301309013002.
51. J.P. Blocki et al. Nuclear asymmetry energy and isovector stiffness within the effective surface approximation. Phys. Rev. C 87 (2013) 044304.
DOI: 10.1103/PhysRevC.87.044304.
52. J.P. Blocki, A.G. Magner, P. Ring. Slope-dependent nuclear-symmetry energy within the effective-surface approximation. Phys. Rev. C 92 (2015) 064311.
DOI: 10.1103/PhysRevC.92.064311.
53. J.S. Rowlinson. Translation of J. D. van der Waals "The thermodynamik theory of capillarity under the hypothesis of a continuous variation of density" J. Stat. Phys. 20 (1979) 197.
DOI: 10.1007/BF01011513.
54. J.S. Rowlinson, B. Widom. Molecular Theory of Capillarity (Clarendon Press, Oxford, 1982).
55. M. Brack, C. Guet, H.-B. Håkansson. Selfconsistent semiclassical description of average nuclear properties a link between microscopic and macroscopic models. Phys. Rep. 123 (1985) 275.
DOI: 10.1016/0370-1573(86)90078-5.
56. M. Brack, R.K. Bhaduri. Semiclassical Physics (Addison-Wesley, 1997; 2nd ed. Westview Press, 2003).
57. J. Lense, H. Thirring. Über den Einfluss der Eigenrotation der Zentralkörper auf die Bewegung der Planeten und Monde nach der Einsteinschen Gravitationstheorie. Phys. Zeit 19 (1918) 156. English translation in Gen. Relativ. Gravit. 16 (1984) 727.
DOI: 10.1007/BF00762913.
58. A. Bohr, B. R. Mottelson. Nuclear Structure. Vol. II: Nuclear Deformations (W.A. Benjamin, New York,

1975).
DOI: 10.1063/1.3037453.

59. D. Vautherin, D.M. Brink. Hartree–Fock calculations with Skyrme's interaction. I. Spherical nuclei. Phys. Rev. C 5 (1972) 626.
DOI: 10.1103/PhysRevC.5.626.
60. T.H.R. Skyrme. CVII. The nuclear surface. Philos. Mag. 1 (1956) 1043.
DOI: 10.1080/14786435608238186.
61. R.C. Barrett, D.F. Jackson. Nuclear Sizes and Structure (Clarendon Press, Oxford, 1977).
62. P. Ring, P. Schuck. The Nuclear Many-Body Problem (Springer-Verlag, Berlin, Heidelberg, New York, 1980).
DOI: 10.1007/978-3-642-61885-4.
63. J.P. Blaizot. Nuclear compressibilities. Phys. Rep. 64 (1980) 171.
DOI: 10.1016/0370-1573(80)90001-0.
64. B. Grammaticos, A. Voros. Semiclassical approximations for nuclear giant resonances. Ann. Phys. 123 (1979) 359; 127 (1980) 153.
DOI: 10.1016/0003-4916(79)90132-4, 10.1016/0003-4916(80)90209-8.
65. H. Krivine, J. Treiner, O. Bohigas. Derivation of a fluid-dynamical Lagrangian and electric giant resonances. Nucl. Phys. A 336 (1980) 155.
DOI: 10.1016/0375-9474(80)90206-8.
66. E. Chabanat, et al. A Skyrme parametrization from subnuclear to neutron star densities. Nucl. Phys. A 627 (1997) 710.
DOI: 10.1016/S0375-9474(97)00596-4.
67. E. Chabanat, et al. A Skyrme parametrization from subnuclear to neutron star densities. II. Nuclei far from stability. Nucl. Phys. A 635 (1998) 231.
DOI: 10.1016/S0375-9474(98)00180-8.
68. W.D. Myers, W.J. Swiatecki. The nuclear droplet model for arbitrary shapes. Ann. Phys. 84 (1974) 186.
DOI: 10.1016/0003-4916(74)90292-8.
69. W.D. Myers, W.J. Swiatecki, C.S. Wang. The surface energy of multi-component systems. Nucl. Phys. A 436 (1985) 185.
DOI: 10.1016/0375-9474(85)90338-4.
70. P. Danielewicz, J. Lee. Symmetry energy in nuclear surface. Int. J. Mod. Phys. E 18 (2009) 892.
DOI: 10.1142/S0218301309012857.
71. M. Centelles, X., et al. Nuclear symmetry energy probed by neutron skin thickness of nuclei. Phys. Rev. Lett. 102 (2009) 122502.
DOI: 10.1103/PhysRevLett.102.122502.
72. M. Centelles, et al. Origin of the neutron skin thickness of 208Pb in nuclear mean-field models. Phys. Rev. C 82 (2010) 054314.

DOI: 10.1103/PhysRevC.82.054314.

73. X. Roca-Maza, M. Centelles, X. Viñas, M. Warda. Neutron skin of 208Pb, nuclear symmetry energy, and the parity radius experiment. Phys. Rev. Lett. 106 (2011) 252501.
DOI: 10.1103/PhysRevLett.106.252501.
74. X. Viñas., et al. Density dependence of the symmetry energy from neutron skin thickness in finite nuclei. Eur. Phys. J. A 50 (2014) 27.
DOI: 10.1140/epja/i2014-14027-8.
75. J. Piekarewicz, M. Centelles. Relativistic mean-field models and the neutron skin thickness of 208Pb. Phys. Rev. C 79 (2009) 054311.
DOI: 10.1103/PhysRevC.79.054311.
76. T. Nikšić, D. Vretenar, P. Ring. Relativistic nuclear energy density functionals: Mean-field and beyond. Prog. Part. Nucl. Phys. 66 (2011) 519.
DOI: 10.1016/j.ppnp.2011.01.055.
77. N. Chamel, P. Haensel. Physics of neutron star crusts. Living Rev. in Relativity 11 (2008) 10.
DOI: 10.12942/lrr-2008-10.
78. B. T. Reed et al. Implications of PREX-2 on the equation of state of neutron-rich matter. Phys. Rev. Lett. 126 (2021) 172503.
DOI: 10.1103/PhysRevLett.126.172503.
79. R. P. Kerr. Gravitational field of a spinning mass as an example of algebraically special metrics. Phys. Rev. Lett. 11 (1963) 237.
DOI: 10.1103/PhysRevLett.11.237.
80. S. A. Teukolsky. The Kerr metric. Class. Quantum Grav. 32 (2015) 124006.
DOI: 10.1088/0264-9381/32/12/124006.
81. S. Schuster, M. Visser. Boyer–Lindquist space-times and beyond: metamaterial analogues for arbitrary space-times. Universe 10 (2024) 159.
DOI: 10.3390/universe10040159.
82. J. B. Hartle. Slowly rotating relativistic stars. I. Equations of structure. The Astrophys. J. 150 (1967) 1005.
DOI: 10.1119/1.1783900.
83. R. H. Boyer, R. W. Lindquist. Maximal analytic extension of the Kerr metric. J. Math. Phys. 8 (1967) 265.
DOI: 10.1063/1.1705193.
84. P. A. Hogan. An interior Kerr solution. Lett Nuovo Cim. 16 (1976) 33.
DOI: 10.1007/BF02746870.
85. A. Krasiński. Ellipsoidal space-times, sources for the Kerr metric. Ann. Phys. 112 (1978) 22.
DOI: 10.1016/0003-4916(78)90079-9.
86. J. L. Friedman, J. R. Ipser, L. Parker. Rapidly rotating neutron star models. Astrophys. J. 304 (1986) 115 (ISSN 0004-637X).

87. N. Stergioulas. Rotating stars in relativity. Living Rev. Relativity (Max Planck Institute for Gravitational Physics, Albert Einstein Institute, Germany, 2003).

88. A. Worley, P. G. Krastev, B.-A. Li. Nuclear constraints on the moments of inertia of neutron stars. Astrophys. J. 685 (2008) 390.
DOI: 10.1086/589823.

89. J. L. Friedman, N. Stergioulas. Rotating Relativistic Stars (Cambridge University Press, Cambridge, New York, 2013).
DOI: 10.1007/s41114-017-0008-x.

90. R. H. Boyer, R. W. Lindquist. A variational principle for a rotating relativistic fluid. Phys. Lett. 20 (1966) 504.
DOI: 10.1016/0031-9163(66)90975-9.

91. A.G. Magner et al. Rotating neutron stars within the macroscopic effective surface approximation. arXiv:2509.13129v1 [gr-qc] 2025, submitted now to the Reports on Progress in Physics (2026).

**А. А. Улеєв[1], О. Г. Магнер[1], С. П. Майданюк[2,3], А. Бонасера[4], Х. Женг[5,*], С. М. Федоткін[1], О. І. Левон[6], У. В. Григор'єв[1], Т. Депастас[4]**

[1] *Інститут ядерних досліджень НАН України, відділ теорії ядра, Київ, Україна*
[2] *Інститут ядерних досліджень НАН України, відділ теорії ядерних процесів, Київ, Україна*
[3] *Інститут сучасної фізики, Китайська академія наук, Хуіжоу, Китай*
[4] *Циклотронний інститут, Техаський університет A&M, Коледж-Стейшн, Техас, США*
[5] *Школа фізики та інформаційних технологій, Шеньсійський педагогічний університет, Сіань, Китай*
[6] *Інститут ядерних досліджень НАН України, відділ фізики важких іонів, Київ, Україна*

*Відповідальний автор: zhengh@snnu.edu.cn

## МАКРОСКОПІЧНІ НАБЛИЖЕННЯ ДО ОБЕРТАННЯ НЕЙТРОННИХ ЗІРОК

Макроскопічна модель нейтронної зірки (НЗ) як ідеальної рідкої краплини поширюється на обертальні системи з малою швидкістю обертання $\omega$ в наближенні ефективної поверхні (ЕП). Враховано градієнтні члени енергії густини НЗ $\mathcal{E}(\rho)$ рівняння стану разом з об'ємними компонентами в головному наближенні за лептодермічним параметром $a/R \ll 1$, де $a$ є товщиною шару ЕП та $R$ – середній радіус НЗ. Для розрахунку адіабатичного моменту інерції (МІ) використовується вираз макроскопічного кутового моменту за малої частоти $\omega$ в рамках метрики Керра у зовнішніх координатах Бойера-Лінгвіста та внутрішніх координатах Хогана. Отримано НЗ МІ $\Theta = \widetilde{\Theta}/\left(1 - \mathcal{G}_{t\varphi}\right)$ через статистично усереднений МІ $\widetilde{\Theta}$ та азимутально-кутову кореляцію $\mathcal{G}_{t\varphi}$ як суму об'ємного та поверхневого компонентів. Показано, що МІ $\Theta$ драматично залежить від ефективного радіусу $R$ через сильну гравітацію та поверхневих ефектів. Знайдено значні додаткові обмеження на радіус $R$ завдяки кореляції $\mathcal{G}_{t\varphi}$ та поверхневим внескам. З цими внесками адіабатична умова при сильній гравітації виконується для багатьох добре відомих НЗ.